\documentclass[conference]{IEEEtran}
\IEEEoverridecommandlockouts
\usepackage{cite}
\usepackage{amsmath,amssymb,amsfonts}
\usepackage{algorithmic}
\usepackage{graphicx}
\usepackage{listings}
\usepackage{textcomp}
\usepackage{xcolor}
\usepackage{comment}

\def\BibTeX{{\rm B\kern-.05em{\sc i\kern-.025em b}\kern-.08em
    T\kern-.1667em\lower.7ex\hbox{E}\kern-.125emX}}

\begin{document}

\title{Efficient techniques to GPU Accelerations of Multi-Shot Quantum Computing Simulations}

\author{\IEEEauthorblockN{1\textsuperscript{st} Jun Doi}
\IEEEauthorblockA{\textit{IBM Quantum} \\
\textit{IBM Research - Tokyo}\\
Tokyo, Japan \\
doichan@jp.ibm.com}
\and
\IEEEauthorblockN{2\textsuperscript{nd} Hiroshi Horii}
\IEEEauthorblockA{\textit{IBM Quantum} \\
\textit{IBM Research - Tokyo}\\
Tokyo, Japan \\
horii@jp.ibm.com}
\and
\IEEEauthorblockN{3\textsuperscript{rd} Christopher Wood}
\IEEEauthorblockA{\textit{IBM Quantum} \\
\textit{IBM Research}\\
Yorktown Heights, United States \\
cjwood@us.ibm.com}
}
\maketitle

\begin{abstract}
Quantum computers are becoming practical for computing numerous applications. However, simulating quantum computing on classical computers is still demanding yet useful because current quantum computers are limited because of computer resources, hardware limits, instability, and noises. Improving quantum computing simulation performance in classical computers will contribute to the development of quantum computers and their algorithms. Quantum computing simulations on classical computers require long performance times, especially for quantum circuits with a large number of qubits or when simulating a large number of shots for noise simulations or circuits with intermediate measures. Graphical processing units (GPU) are suitable to accelerate quantum computer simulations by exploiting their computational power and high bandwidth memory and they have a large advantage in simulating relatively larger qubits circuits. However, GPUs are inefficient at simulating multi-shots runs with noises because the randomness prevents highly parallelization. In addition, GPUs have a disadvantage in simulating circuits with a small number of qubits because of the large overheads in GPU kernel execution. In this paper, we introduce optimization techniques for multi-shot simulations on GPUs. We gather multiple shots of simulations into a single GPU kernel execution to reduce overheads by scheduling randomness caused by noises. In addition, we introduce shot-branching that reduces calculations and memory usage for multi-shot simulations. By using these techniques, we speed up x10 from previous implementations.

\end{abstract}

\begin{IEEEkeywords}
quantum computing, simulation, noise simulation, GPU, Qiskit
\end{IEEEkeywords}

\section{Introduction}
Since quantum computers have become practical for calculating actual applications, quantum computing simulations on classical computers still have a high demand for the development of quantum algorithms, software, and hardware. Quantum computing simulations on classical computers are very useful to reproduce the environments of actual quantum computers, such as noise behaviors, and can also produce an ideal environment without noise or break the limitations of circuit size and the number of qubits.

However, simulating quantum computing on classical computers requires huge computational power and memory space. To simulate a \verb+n-qubits+ circuit, it requires $2^n$ floating point complex numbers to store probability amplitudes of state vectors and $O(2^n)$ computations. For example, a \verb+30-qubits+ circuit simulation requires at least \verb+16GB+ of memory in double precision. Therefore, simulating circuits with a large number of qubits requires highly parallel computers to store big data and accelerate computations.

Another aspect of a high computational requirement for quantum computing simulations is to simulate a large number of shots of execution on the same quantum circuit because many quantum algorithms reduce the results statistically to finalize the answer. On the simulator, if there is no noise and the measurements to all the qubits are placed at the end of the circuit, we can apply the sampling measures technique to obtain the measured results for all shots at once. However, if there are noises or branch operations in the middle of the circuit, we have to run all the shots independently. Therefore, multi-shot simulations also have a large demand running on high-performance computers.

Because quantum computing simulations are memory bandwidth-bound applications, a graphical processing unit (GPU) is a powerful tool to accelerate quantum computing simulations. As a GPU has high bandwidth memory with a large number of threads, it is suitable to simulate quantum circuits with a large number of qubits. However, a GPU has less memory space than that of host computers, which limits the number of qubits. Therefore, a number of techniques to distribute probability amplitudes over multiple GPUs or GPUs on clusters are required to run circuits with a large number of qubits.

In addition, a GPU is not good at accelerating small problem sizes because of overheads to launch kernel programs or non-uniform calculations caused by the randomness of multi-shot simulations with noise sampling or that triggered by intermediate measures. A number of techniques to optimize multi-shot simulations on GPUs are demanding.

In this paper, we propose two techniques: batch-shots execution simulates multiple shots in a single GPU kernel program to decrease kernel launch overheads, and shot-branching simulates a single state shared with multiple shots and branch states when randomness occurs.

To handle the randomness of the noises or measurements in batch-shots execution, we propose a technique to synchronize between shots. We implemented an optimized multi-shot simulation technique in an open-source quantum computing simulation platform, Qiskit Aer \cite{qiskit-aer}. Qiskit Aer supports accelerating simulations using GPUs and parallelizes large qubits simulations or distributes multi-shot simulations using multiple GPUs or multiple compute nodes. However, multi-shot simulations for small qubits simulations on a GPU were slower than simulating on a host central processing unit (CPU) because of its large overheads. This is our motivation to improve multi-shot simulations on GPUs.

In shot-branching, we generate a tree structure of states on the fly while simulating a quantum circuit. In this structure, the states of the qubits are shared with multi-shots that have the same state. From the root of the tree, there is only a single state shared with all the shots, and state will be branched into several new states when randomness occurs such as noises, measures, initializes, and resets. So we can decrease calculations and memory requirements if number of states generated by shot-branching is smaller than number of shots. This technique is effective for noise simulations, because the error rate is usually small and there will be small number of branches in most cases. This technique is not only for GPUs but also applicable to simulation on CPU.

In section II we mention a number of efforts on prior work for the quantum computing simulations. We introduce overview of Qiskit Aer and the issues in the noise simulations using GPUs on Qiskit Aer in Section III. Then we introduce our optimization techniques for multi-shot simulations in section IV and we compare and evaluate noise simulations in section V.

\section{Related Work}
Because quantum computing simulation on the classical computers requires high memory bandwidth and high performance computing, there have been a lot of efforts to accelerate quantum computing simulators by exploiting GPUs. Qsim \cite{qsim} is a simulator backend of Cirq \cite{cirq} platform that accelerates simulations by using SIMD and OpenMP threads on CPU and also GPU. Qulacs \cite{Suzuki2021qulacsfast} is another simulator backend can be used for Cirq interface that support GPU acceleration. Qibo \cite{efthymiou2021qibo} also accelerates simulation using GPU and multiple GPUs. QuEST \cite{jones2018quest} and Qiskit Aer are also simulators supporting GPU.

Also there are efforts on simulating large scale circuits by parallelizing and distributing state vectors on parallel processes \cite{li2017quantum,haner20170,jones2018quest,smelyanskiy2016qhipster,DeRaedt2019,wu2019,doi2019,doi2020,Liu2021}. 

Most of the simulators support simulating noises on the quantum circuits. \cite{isakov2021} introduces optimization technique of noise simulation with Kraus operator on qsim and Cirq. A density matrix method \cite{benenti2007} is another common technique to simulate noise simulation and implemented in qsim, Amazon Braket \cite{amazon_braket} and Qiskit Aer. A density matrix method can simulate multiple shots of simulation with noises by a single shot of simulation run, but it requires $O(4^N)$ memory space and calculations. So density matrix method is only applicable for small qubits simulations.

Because quantum computing simulation demands large data and bandwidth, most of prior work is focused on data-wise parallelization and distribution of large scale state vectors. Another aspect of improvement of quantum computing simulation is shot-wise optimization and parallelization to speed up noisy circuit simulation. Because a number of of the shots has the same probability amplitudes during the simulation, calculations and probability amplitudes can be shared and reused among a number of shots that decreases calculation and memory space for multi-shot simulations. \cite{Zulehner2017,Li2019,Wang2022} introduces technique to decrease calculations of multi-shot simulations by using tree structures to record state to handle branches of shots recursively. Our shot-branching technique is same idea but we do not record states for recursive execution but branched states are parallely simulated.

\section{Quantum Computing Simulation on Qiskit Aer}
\subsection{Qiskit Aer Overview}
Qiskit Aer \cite{qiskit-aer} is one of the components of Qiskit \cite{qiskit}, an open-source quantum computing platform. And Qiskit Aer provides a simulator backends on the classical computers that accepts the same interface to the actual quantum computer to execute quantum circuits. Qiskit Aer has various simulation methods such as state vector, density matrix, stabilizer and matrix product state methods and various noise models that reproduce behaviors of noisy environment of the actual quantum computers. 

\subsection{Multi-shots and Noise Simulations in Qiskit Aer}
Quantum circuits simulations on Qiskit Aer are usually multi-shot simulations because most quantum algorithms statistically reduce measured values to obtain a number of results. A multi-shot simulation runs a single quantum circuit and returns statistical results in a bit count array. Qiskit Aer calculates each multi-shot in separate states when applying noise models or when there are conditional gate operations with intermediate measurements. 

Qiskit Aer has various types of noise models that simulate actual noisy quantum computer behaviors. In addition to using a number of predefined noise models, users can add customized ones. Quantum errors are noises caused by gate operations that negatively affect probability amplitudes, and read out errors are noises that affect measured values. In this paper, we focus on quantum errors, especially for the two common noise models defined in Qiskit Aer: Pauli and Kraus.

\subsubsection{Pauli Noise Model}
The Pauli noise is represented as combinations of bit-flip and phase-flip errors by Pauli gates (X, Y, and Z gates). It is modeled by providing a list of possible Pauli gate error combinations. Each combination is sampled by comparing it to a random number and has a probability value related to the error ratio. If the given random number is lower than the probability, the combination is sampled as a noise. Table \ref{table_pauli_noise_example} shows a simple example of the Pauli noise model with a 1\% bit-flip error represented by an X gate. However in most cases, an ID gate is sampled, which means no error occurred.

\begin{table}[!t] \renewcommand{\arraystretch}{1.3}
\caption{Example of Pauli noise model that has a 1\% bit-flip error}
\label{table_pauli_noise_example} \centering 
\begin{tabular}{c|c}
\hline \hline \bfseries Probabilities & \bfseries Pauli gates\\
\hline 0.99 & ID\\
\hline 1.00 & X\\
\hline 
\end{tabular} \end{table}

The following Python script shows an example of how to simulate a quantum Foulier transform (QFT) circuit with the Pauli noise model on Qiskit Aer. In this example, 1\% of Pauli errors are applied to \verb+1-qubit+ gates U1, U2, U3, X, H, and P, and \verb+2-qubits+ gate CX.

\begin{lstlisting}[language=python]
from qiskit import *
from qiskit.circuit.library import *
from qiskit.providers.aer import *
from qiskit.providers.aer.noise import *
noise_model = NoiseModel()
error1 = depolarizing_error(0.01, 1)
error2 = depolarizing_error(0.01, 2)
noise_model.add_all_qubit_quantum_error(error1, ['u1', 'u2', 'u3','x','h','p'])
noise_model.add_all_qubit_quantum_error(error2, ['cx'])
sim = AerSimulator(method='statevector', noise_model = noise_model)
circuit = transpile(QFT(10),basis_gates=noise_model.basis_gates)
circuit.measure_all()
execute(circuit,sim,shots=1000,noise_model=noise_model)
\end{lstlisting}

\subsubsection{Kraus Noise Model}
The Kraus noise model is represented as a Kraus operator. A Kraus operator contains a list of Kraus matrices. Any kind of noise model can be represented by the matrices, so Pauli noises can also be modeled as a Kraus noise model. Similarly to the Pauli noise model, a random number is used to select a matrix to be multiplied with the states, but a Kraus operator takes the expected norm of probability amplitudes after matrix multiplication for a comparison. The following pseudo code is an implementation of a Kraus operator.

\begin{lstlisting}
rnd = random()
prob = 0
for matrix in Kraus_noise
    p = expval_matrix(state, matrix)
    prob += p
    if rnd < prob then
        apply_matrix(state, matrix/p)
        break
\end{lstlisting}

\subsection{GPU Acceleration and Issues in Qiskit Aer}

\subsubsection{Overview of GPU Support in Qiskit Aer}
Qiskit Aer can accelerate state vectors, unitary matricies, and density matrix methods using GPUs. Multiple GPUs can be used to parallelize simulations; probability amplitudes of states are distributed to multiple GPUs if there is insufficient memory space to store all of them on a single GPU, or multi-shots are distributed to multiple GPUs for circuits with a smaller number of qubits. In addition, Qiskit Aer supports the message passing interface (MPI) to distribute and parallelize on GPU clusters.

Qiskit Aer uses the Thrust \cite{Thrust} C++ library to manage probability amplitudes stored on GPUs and implement GPU kernel programs of quantum gate operations. Thrust is a template class library that has a vector template class to manage GPU memory in a similar form to a \verb+std::vector+ template vector class. Kernel programs can be written using lambda functions or functors to make them readable for developers who lack GPU programming knowledge, which is important for open-source projects. In addition, kernel programs written using the Thrust library can be executed on both the GPU and host computer, which is useful for debugging and testing. 

Qiskit Aer also support NVIDIA's cuQuantum \cite{cuquantum} software development kit (SDK) to accelerate quantum computing simulations. Qiskit Aer exploits cuStateVec \cite{custatevec} APIs, one of the components of cuQuantum SDK, to accelerate state vectors, unitaries, and the density matricies.

\subsubsection{Performance Issues in GPU Acceleration}
\begin{figure}[!t] \centering \includegraphics[width=2.5in]{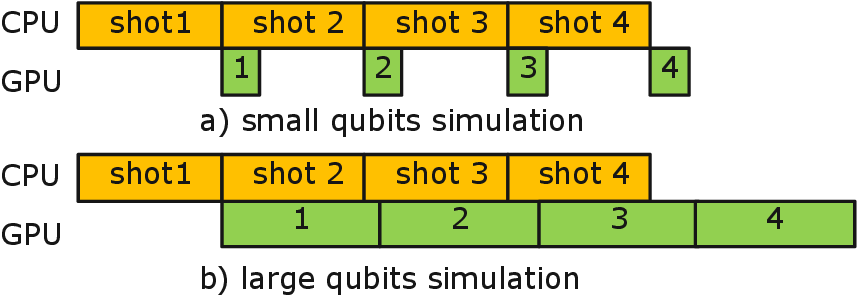} 
\caption{Comparison of GPU overheads loaded on host CPU for multi-shot simulation. a) A small number of qubits have relatively large overheads compared with b) a large number of qubits, where the overheads can be ignored.}
\label{fig_GPU_overhead} 
\end{figure}

A GPU is a powerful computer to accelerate uniform and large scale calculations by utilizing its high bandwidth memory and large number of threads. This means we have to prepare large and uniform problems to obtain the full power of the GPU. A small problem size also causes performance regression because of the overheads loaded on the host computer that contain kernel launch overheads and the memory transfer between the host and GPU. Fig. \ref{fig_GPU_overhead} illustrates the relative impact of GPU overheads loaded on the host CPU compared with the problem size. If the problem size is too small, overheads dominate the execution time and we cannot obtain sufficient GPU acceleration. On the other hand, if the problem size is sufficiently large, we can ignore overheads and obtain sufficient acceleration.

\begin{figure}[!t] \centering \includegraphics[width=2.5in]{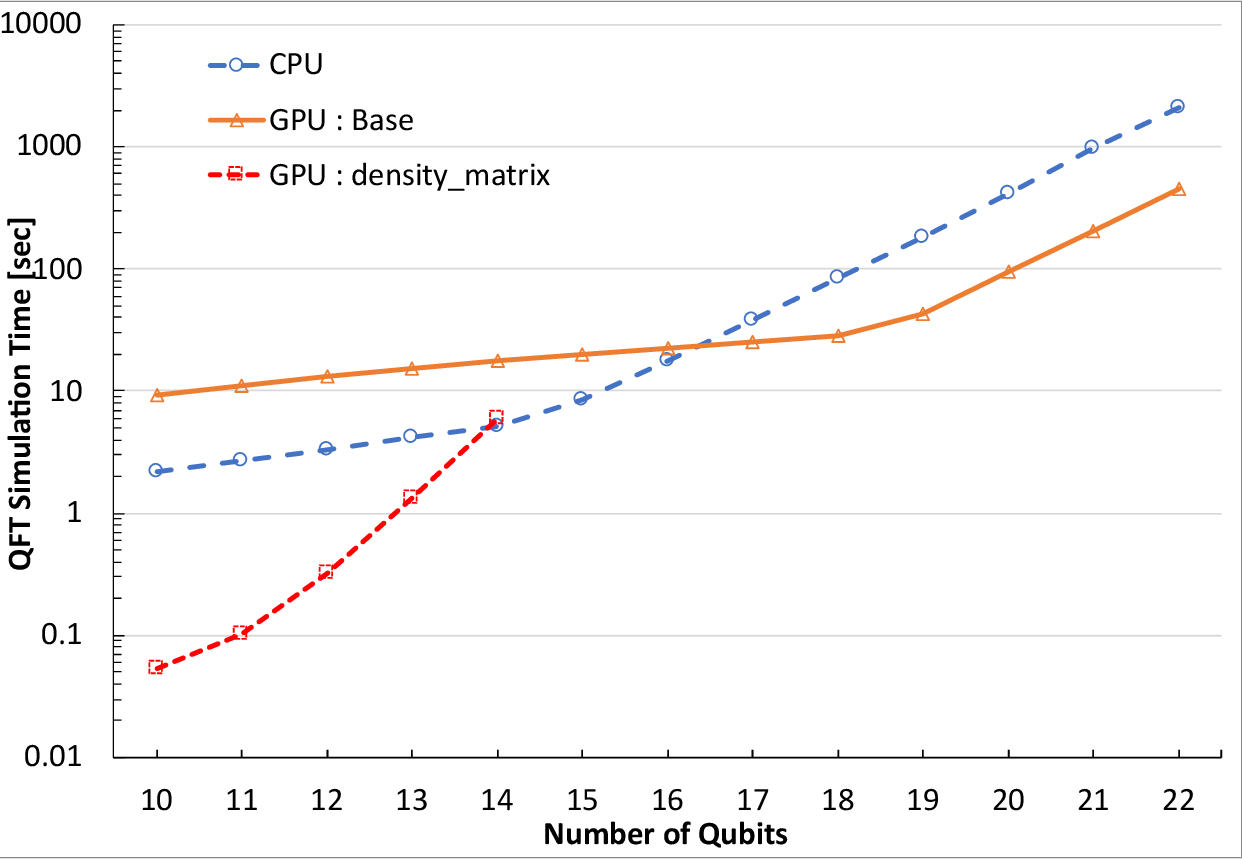} 
\caption{Simulation time comparison of QFT circuit with Pauli noise (4000 shots, error rate = 0.01) on Qiskit Aer running on an IBM Power System AC922 with a single NVIDIA Tesla V100 (16GB)}
\label{fig_QFT_pauli_init} 
\end{figure}

Fig. \ref{fig_QFT_pauli_init} shows a comparison of simulation time of a QFT circuit with 1\% of Pauli noise running on Qiskit Aer. Because of the GPU overheads, much time is required to simulate on the GPU than the CPU for smaller qubits. In addition, Fig. \ref{fig_QFT_pauli_init} plots the simulation time using the density matrix method on GPU. It shows a large increment of time versus the number of qubits compared with a multi-shots state vector method simulation, and we can only simulate up to \verb+14-qubits+ on a single GPU because of its memory limit. Therefore, the density matrix method is only applicable to noise simulations with a small number of qubits.

Multi-shot simulations have non-uniform operations varied for each shot, so it is not easy to apply shot-wise parallelization on a GPU. If the number of qubits is suffiently large, each shot can be parallelized independently with a large number of threads on the GPU because each shot has sufficient parallelism. However, simulating a smaller number of qubits has less parallelism and the overheads are relatively very large, which decrease performance. Therefore, shot-wise parallelization is necessary for a smaller number of qubits. In the next section, we describe how we apply shot-wise parallelization and accelerate multi-shot simulations for a smaller number of qubits.

\section{Acceleration of Multi-shot Simulations}
\subsection{Implementing Batch-Shots Technique}
\subsubsection{Scheduling and Batching Multi-shot Simulation}
\begin{figure}[!t] \centering \includegraphics[width=2.5in]{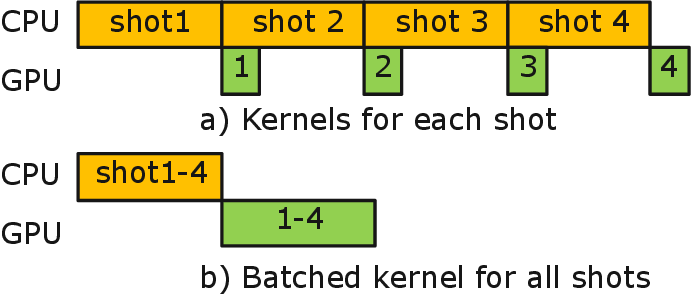} 
\caption{Batched execution of multi-shots in a single kernel. a) Calling the kernel for each shot loads large GPU overheads to host CPU, versus b) calling only one overhead for batched kernel execution. }
\label{fig_batched_kernel} 
\end{figure}

Shot-wise parallelization can be exploited when the same gate operation is applied to all shots. Since the original quantum circuit is identical and branched by noises or measurements, we can synchronize the timing so that the original gates are applied to all shots at the same time. We apply a batched execution model to calculate the gate operations for multi-shots in a single GPU kernel program. Fig. \ref{fig_batched_kernel} illustrates the effect of batched kernel execution. In this example, four shots takes four large overheads, but when we batch the four shots in a single kernel program, the calculation time is increased four times but it only takes one large overhead, which dramatically decreases the total execution time. 

The remaining gates are those added by noise sampling and measurements and conditional ones referring to the measured results. In the following subsections, we describe a number of techniques to synchronize and execute these gates in a batched single kernel program. In this strategy, these batched kernel programs sometimes perform unnecessary calculations for a number of shots, but the unnecessary calculation time will be much smaller than the GPU overheads for a smaller number of qubits.

\subsubsection{Data Structure for Batched Multi-Shot Kernel Execution}
\begin{figure}[!t] \centering \includegraphics[width=2.5in]{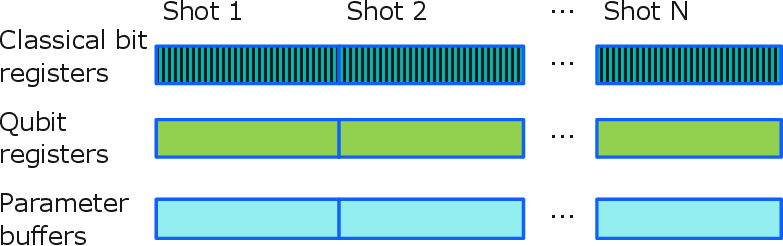} 
\caption{Data structure for multi-shot simulation on a GPU. There are three vectors of data: classical bit register to store conditions, qubit register to store probability amplitudes, and parameter buffers to store data used for gate operations for each shot.}
\label{fig_multi_shots_data_structure} 
\end{figure}

To apply batched multi-shot execution on a GPU, multiple quantum states are stored in the GPU memory. To update the probability amplitudes of each state in a single kernel program, the amplitudes are stored sequentially in a single array to enable us to reuse the same kernel programs applied for a single shot execution. Fig. \ref{fig_multi_shots_data_structure} illustrates the data arrays stored in the GPU for batched multi-shot execution. The probability amplitudes for each shot are stored in an array of qubit registers. An array of classical bit registers is used to store conditions updated by measurements or branch operations, and these bits are referred to as mask shots if the conditional gate operations can be applied to the shot. Parameter buffers are used to store the parameters used in the gate operations such as matricies and an array of qubits containing target or control bits. Most gate operations uses the same parameters, but noise operations take different parameters for each shot.

In a batched single kernel program, the index of each shot can be calculated by the following equation where $i_{shot}$ is the index of a shot, $i$ is an iteration count of the kernel program, $n_q$ is the number of qubits of a circuit, and $n_g$ is the number of qubits of the gate operation. By using this index, we can refer to classical bits and parameters for each shot.
$$i_{shot}=\frac{i}{2^{n_q-n_g}}$$

\subsubsection{Batch Optimization for Pauli Noise}
\begin{figure}[!t] \centering \includegraphics[width=2.5in]{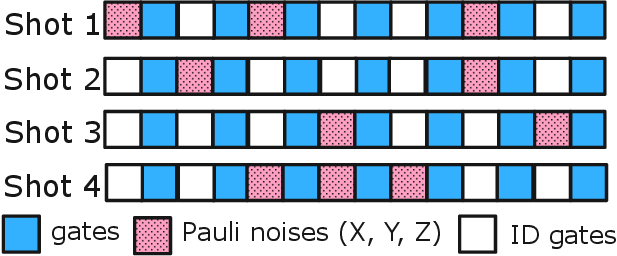} 
\caption{Runtime noise sampling of Pauli noise for each shot. ID gates are added for shots if no noise is sampled to apply Pauli noise in a batched kernel.}
\label{fig_noise_schedule} 
\end{figure}

To calculate the gates added by Pauli noise in a single kernel, we sample Pauli noise at runtime for each shot. Fig. \ref{fig_noise_schedule} illustrates runtime Pauli noise sampling to synchronize all shots. This enables batched kernel programs to be applied, enabling the calculation of the original gates at the same time. Pauli noise is sampled as combinations of Pauli gates in an array for each shot, and an ID gate is sampled if no noise occurred in a shot. Therefore, if ID gates are sampled for all shots, not action is taken for the noise.

Combinations of Pauli gates are represented as bit-flip by X and Y gates, rotation by Y gates, and phase-flip by Z gates. In the state vector simulation, a bit-flip is performed by swapping two probability amplitudes indexed by applying the exclusive-or of amplitude indicies and mask bits calculated by the target bits. A rotation is performed by multiplying a phase calculated by counting the number of Y gates. A phase-flip is performed by referring to the pop count of an index masked by Z target bits. Therefore, we need only four parameters for each shot to apply Pauli noise in a batched single kernel program as shown in the following pseudo code; the maximum number of target bits for the X and Y gates (\verb+x_max+), the bit mask of the target bit for the X gates (\verb+x_mask+), the number of Y gates (\verb+num_y+), and the bit mask of the target bit for the Z gates (\verb+z_mask+). These four parameters are stored in the parameter buffers on the GPU before launching the kernel program. 

\begin{lstlisting}
mask_l = (1 << x_max) - 1
mask_u = not((1 << (x_max+1))-1)
num_y = mod(num_y,4)
if num_y == 0
  phase = 1.0
else num_y == 1
  phase = (0,-1)
else num_y == 2
  phase = (-1,0)
else
  phase = (0,1)
for i < state_vector_size/2
  if x_mask == 0
    i0 = i*2
    i1 = i0 + 1
  else
    i0 = ((1<<1)&mask_u)|(i&mask_l)
    i1 = i0 xor mask_x
  swap(state[i0],state[i1])
  state[i0] = phase*state[i0]
  state[i1] = phase*state[i1]
  if mod(popcount(i0 & z_mask),2) == 1
    state[i0] = -state[i0]
  if mod(popcount(i1 & z_mask),2) == 1
    state[i1] = -state[i1]
\end{lstlisting}

\subsubsection{Batch Optimization for Kraus Noise}
\begin{figure}[!t] \centering \includegraphics[width=2.5in]{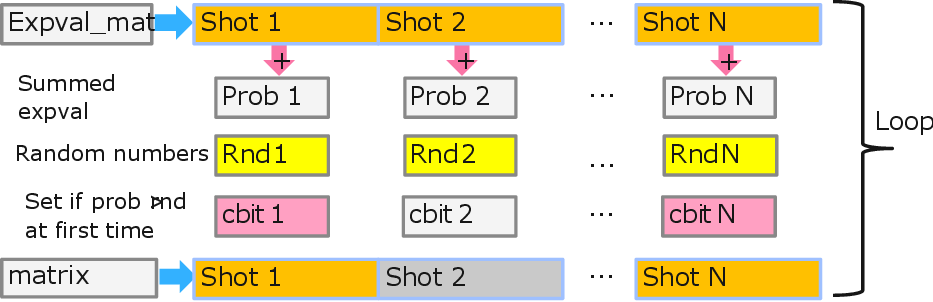} 
\caption{Loop of the batched Kraus operator applied to multi-shots on a GPU. The loop iterates for all possible matrices even if a number of of the shots have completed the multiplication.}
\label{fig_kraus_loop} 
\end{figure}

Kraus noise is applied in a loop for candidate matrices to be multiplied by comparing an expectation value of matrix multiplication with a random probability. Because large overheads are required to return expectation values from the GPU to the host and comparing them on the host, we perform all procedures in the loop on the GPU. In addition, it is more effective to avoid synchronizing the host and GPU. Fig. \ref{fig_kraus_loop} illustrates the loop of batched kernels program for the Kraus noise operator. Since the host CPU does not know the expectation values, we have to run all iterations to the end, even if all shots are already multiplied at early iteration counts. This is trade-off between large GPU overheads and unnecessary loop iterations.

To select and apply one matrix in the candidate list, we set a classical bit if the sum of an expectation value for a shot becomes greater than the random probability the first time, then a matrix is multiplied only with a shot whose classical bit is set. The classical bit is unset after the multiplication.

\subsubsection{Batch Optimization for Measurement and Reset Operation}
We also optimize measure and rest operations to simulate all shots in a batch for circuits with noise models or those with intermediate measurements and branch operations. The measure operation should also be applied independently to each shot because of randomness. The reset operation is applied following the measure operation by referring to the measured bit value, making it also shot dependent.

The following pseudo code shows the measure and reset operation algorithms. The operation enables measuring probabilities for multiple qubits at a time and returns $2^{qubits}$ probabilities. The algorithm looks very similar to the Kraus operator, so we can apply a similar batch optimization to the Kraus operator shown in Fig. \ref{fig_kraus_loop}. First, the sum of the probabilities are calculated by reducing the probability amplitudes whose address is masked by the bit\_mask for each shot. Then, random numbers and the total probabilities sum are compared to evaluate if we can apply the reset operation for each shot and the measured bit is also written in a classical bit.

\begin{lstlisting}
rnd = random()
sum = 0
qubits = (list of target qubits to be measured)
nq = sizeof(qubits)
dim = 2^nq
for i in dim
    bit_mask= 0
    for j in nq
       if ((i >> j) & 1) == 1 then
           bit_mask |= qubits(j)
    prob(i) = probability(bit_mask)
    sum += prob(i)
    if rnd < sum at first time then
        reset(bit_mask)
return prob
\end{lstlisting}

If there are branch and conditional gate operations after the measure operations, these are also shot dependent. The branch gates set the classical bit by referring to the returned classical bits of the measure operation, and the conditional gates are applied only to shots whose classical bit is set.

\subsection{Implementing Shot-Branching Technique}

\subsubsection{Overview of Shot-Branching}
\begin{figure}[!t] \centering \includegraphics[width=2.5in]{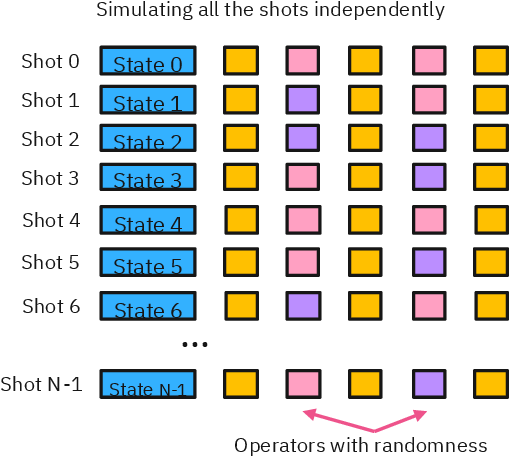} 
\caption{Naive implementation of multi-shot simulation. Each shot is simulated independently in independent quantum states.}
\label{fig_naive_multi_shots} 
\end{figure}

Multi-shot quantum computing simulations should be executed independently for each shot because the quantum state of each shot will be varied because of operations with randomness, for example measure operation will return 0 or 1. Noises simulations should also be executed independently for each shot because noise will be applied randomly to each shot. Fig. \ref{fig_naive_multi_shots} shows an example of quantum operation sequences in a multi-shot simulation with operations involving randomness. Although a number of shots may ultimately have the same quantum state, this is waste of computational power and memory space. Shot-branching is a technique to decrease redundancies of quantum states among shots. This technique is not only designed for GPUs, but also effective for simulating on host computers.

\begin{figure}[!t] \centering \includegraphics[width=2.5in]{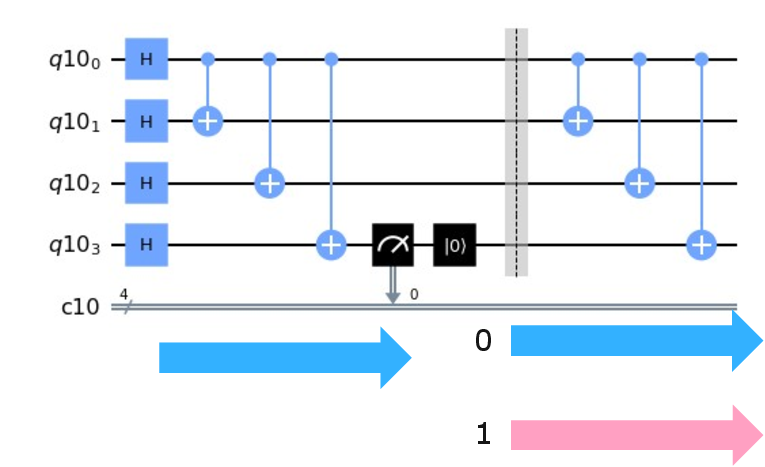} 
\caption{Eexample of quantum circuit with intermediate measurements. Shots are branched depending on the measured bit in two types of quantum state.}
\label{fig_intermediate_measure} 
\end{figure}

Fig. \ref{fig_intermediate_measure} illustrates an example of a quantum circuit that includes an intermediate measurement. The quantum state for all shots is identical prior to the measurement; however, after the measurement, the quantum state will vary depending on the measurement result. In this example, only two final quantum states are possible, but the naive multi-shot simulation calculates for all shots, which is unnecessary.

\subsubsection{Shot-Branching Implementation}
\begin{figure}[!t] \centering \includegraphics[width=2.5in]{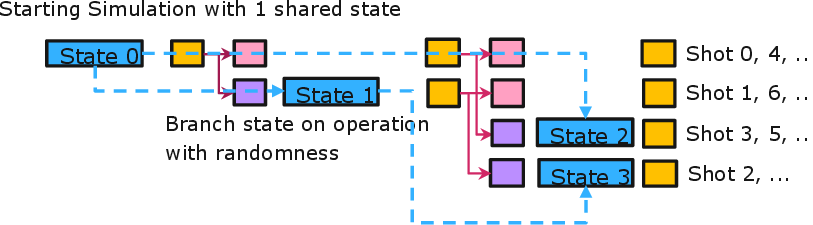} 
\caption{Implementation of the on-the-fly shot-branching. A single quantum state is shared with all shots at first, and the quantum state will be branched and copied when an operation with randomness occurs. Branched states are calculated parallelly if there is sufficient memory to store the branched quantum states.}
\label{fig_shot_branching} 
\end{figure}

The simulation with shot-branching begins with a shared quantum state, where all shots are mapped to the first state. As with the conventional simulation, the state is updated by the sequence of quantum operations. When a quantum gate with randomness is simulated, the quantum state is branched into a number of new states. Fig. \ref{fig_shot_branching} illustrates how the first state is branched into a number of states by the operations with randomness in the same sequence of quantum operations shown in Fig. \ref{fig_naive_multi_shots}. In Fig. \ref{fig_shot_branching}, \verb+state 0+ is branched into \verb+state 0+ and \verb+state 1+ at the first operation with randomness. The quantum state of \verb+state 0+ is copied to \verb+state 1+ and updated independently. Each shot is evaluated and stored in the list of the newly created state. The branched quantum states are then simulated independently, and this procedure is repeated whenever operations with randomness occur.

While branching states, if there is insufficient memory space to allocate a new state, we store the shots in a waiting shots list and continue the simulation on the existing states to the end. Then, we restart the simulation for the shots in the waiting list from the beginning with a new single state. We continue this until waiting shots remain.

Finally, we store the same results to all shots in each state's list. We cannot apply the sampling technique to obtain the final measure for the conventional multi-shot simulation, but we can apply it to the shot-branching technique, which can decrease simulation time if there are still multiple shots remaining in a state's list.

\subsection{Distribution of Multi-shot simulations}
Shot-wise parallelization can be extended to multiple GPUs and compute nodes in a cluster. Since each shot can be calculated independently, shot-wise parallelization is embarrassingly parallel, and communication is only needed to collect results. Shots can be easily distributed to each GPU or node in the cluster. Additionally, shot-branching can be applied to the distributed shots on each simulation location.

\section{Performance Evaluation of Multi-shot Simulations}
\subsection{Computer Environments for the Performance Evaluation}
We used a GPU cluster machine to evaluate the proposed optimization technique for multi-shot simulations implemented on Qiskit Aer. The cluster machine is composed of a IBM Power System AC922 \cite{caldeira2018ibm}, which has six NVIDIA Tesla V100 GPUs \cite{tesla-v100} on each compute node. The detailed specifications of the cluster is shown in Table \ref{table_eval_env}. There are two Power 9 CPU sockets on the node and each socket has three NVIDIA Tesla V100 GPUs. The CPU and GPUs are connected via NVLink2 \cite{nvlink}.

As there is no binary distribution of Qiskit Aer for Linux on Power, we built Qiskit Aer from the source code. In addition, Qiskit Aer needs to be installed from the source code to distribute multi-shots using MPI.

\begin{table}[!t] \renewcommand{\arraystretch}{1.3}
\caption{Computer environment for performance evaluation of noise simulations}
\label{table_eval_env} \centering 
\begin{tabular}{l|l}
\hline 
\hline 
Compute node & IBM Power System AC922 \\\hline
CPU & POWER9 \\\hline
Number of sockets per node & 2 \\\hline
Number of cores per socket & 21 \\\hline
CPU memory size & 512 GB \\\hline
GPU & NVIDIA Tesla V100 \\\hline
Number of GPUs per node & 6 \\\hline
GPU memory size & 16GB \\\hline
Interconnect & Infiniband EDR \\\hline
OS & Red Hat Enterprise Linux Server 7.6 \\\hline
Compiler & GCC 8.3.0 \\\hline
CUDA Toolkit & CUDA 10.1 \\\hline
MPI & IBM Spectrum MPI 10.3.1 \\\hline
\end{tabular} \end{table}

\subsection{Single GPU Performance Evaluation}
First, we compare the simulation time of the QFT circuit with Pauli and Kraus noise models on the GPU and host CPU. In this evaluation, we run 4,000 shots and set the error ratio to 1\%. We also compare the simulation times using the density matrix method on the GPU.

\begin{figure}[!t] \centering \includegraphics[width=2.5in]{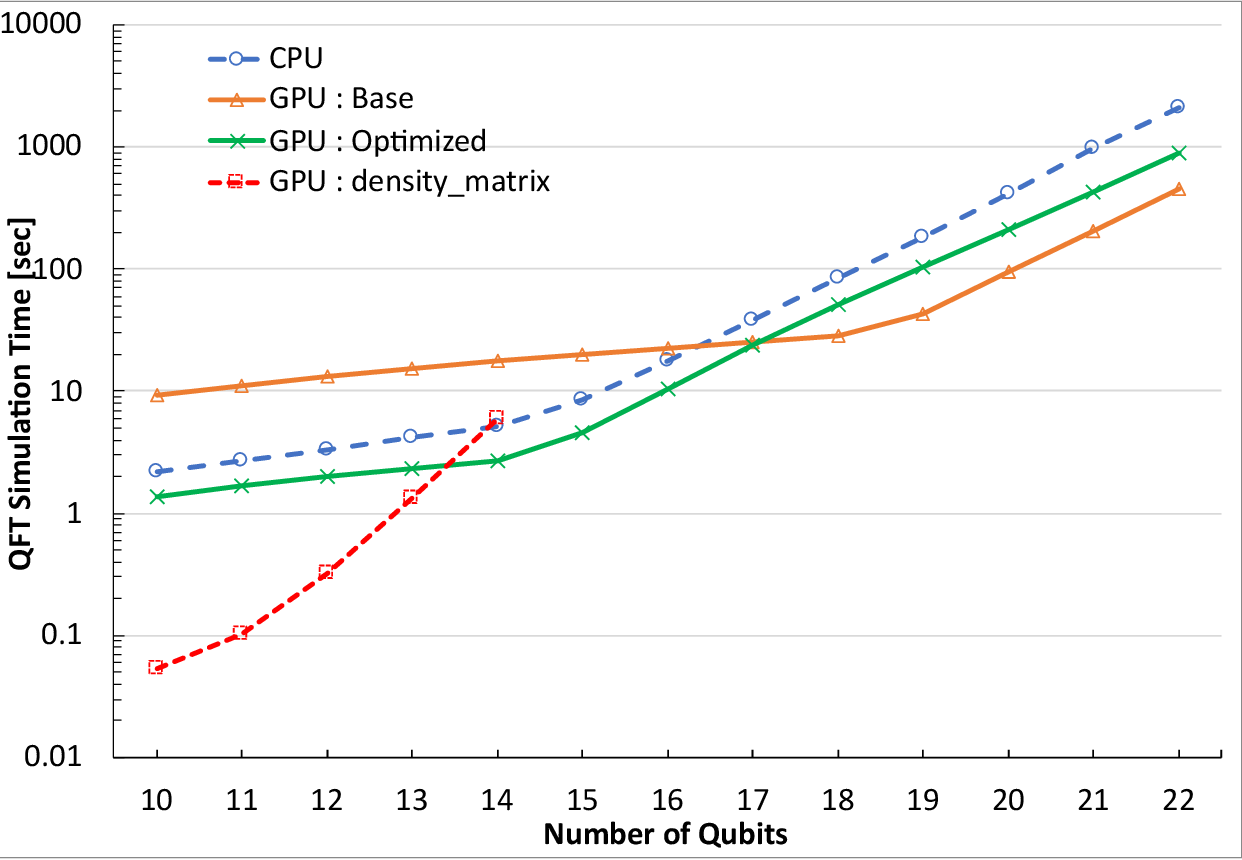} 
\caption{Simulation time comparison of 4,000 shots of QFT circuit with 1\% Pauli noise on Qiskit Aer using a single GPU. (IBM Power System AC922 with 1 NVIDIA Tesla V100.)}
\label{fig_single_GPU_pauli} 
\end{figure}

As shown in Fig. \ref{fig_single_GPU_pauli}, we compared the simulation time of the QFT cicuit with 1\% Pauli noise. In this graph, GPU : Base shows the simulation time using the conventional implementation and GPU : optimized shows that using the batched multi-shot optimization proposed in this paper. The conventional implementation on the GPU was slower than that running on the CPU for a smaller number of qubits because of the large GPU overheads, but the performance was improved by applying the technique proposed in this paper. We increased the performance speed up to about ten times. For a large number of qubits, the conventional implementation is faster because the proposed technique calculates unnecessary ID gates for a number of shots, an noise on a number of shots occurs with a very small probability. The density matrix method has an advantage for Pauli noise simulation with a small number of qubits and a large number of shots.

\begin{figure}[!t] \centering \includegraphics[width=2.5in]{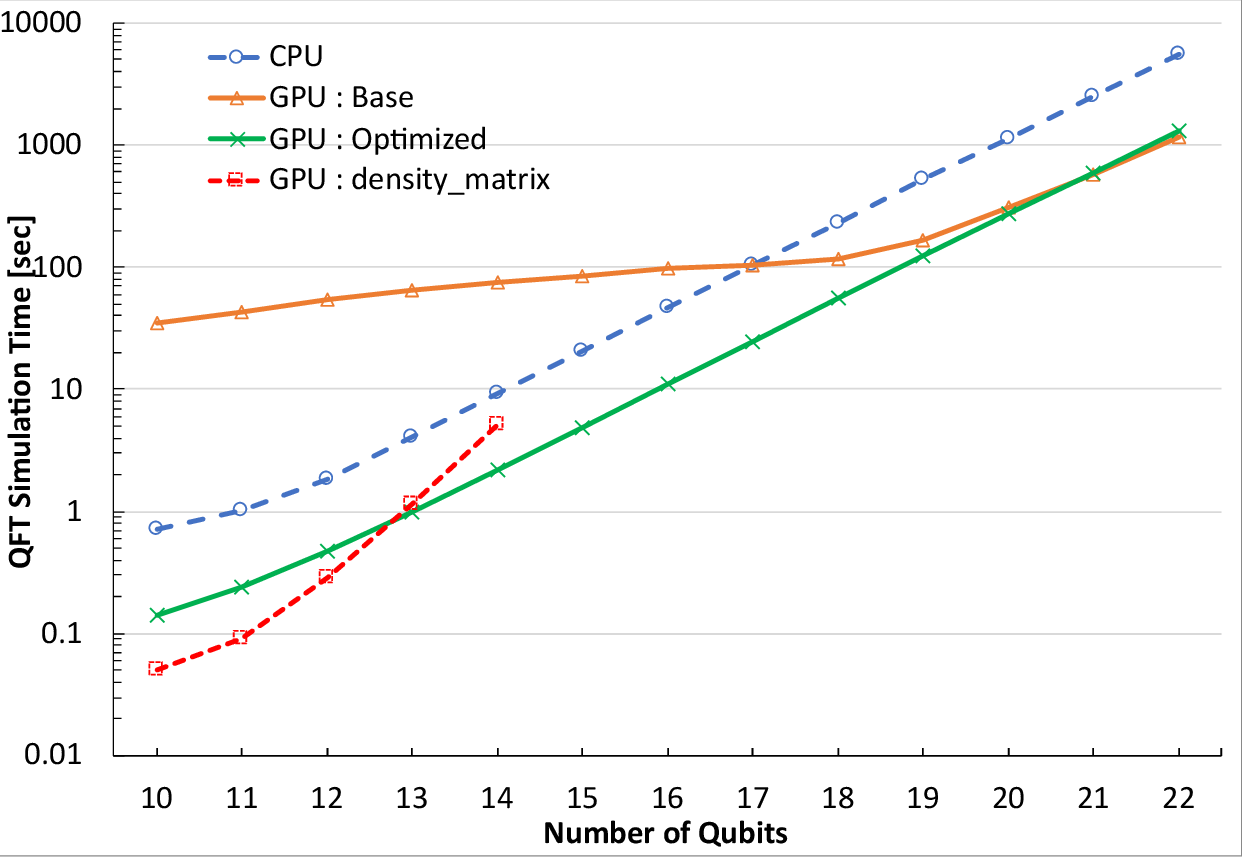} 
\caption{Simulation time comparison of 4,000 shots of QFT circuit with 1\% Kraus noise on Qiskit Aer using a single GPU. (IBM Power System AC922 with 1 NVIDIA Tesla V100.)}
\label{fig_single_GPU_kraus} 
\end{figure}

As shown in Fig. \ref{fig_single_GPU_kraus}, we compared the simulation time of the QFT circuit with 1\% Kraus noise. The conventional implementation was much worse than the situation in the Pauli noise simulation, but we improved the performance largely by applying the batched multi-shot optimization proposed in this paper. We increased the performance speed about ten to hundred times depending on the number of qubits. The increase in speed is almost saturated for a large number of qubits. To obtain the advantage in the density matrix method, we require more shots for a Kraus noise simulation.

\begin{figure}[!t] \centering \includegraphics[width=2.5in]{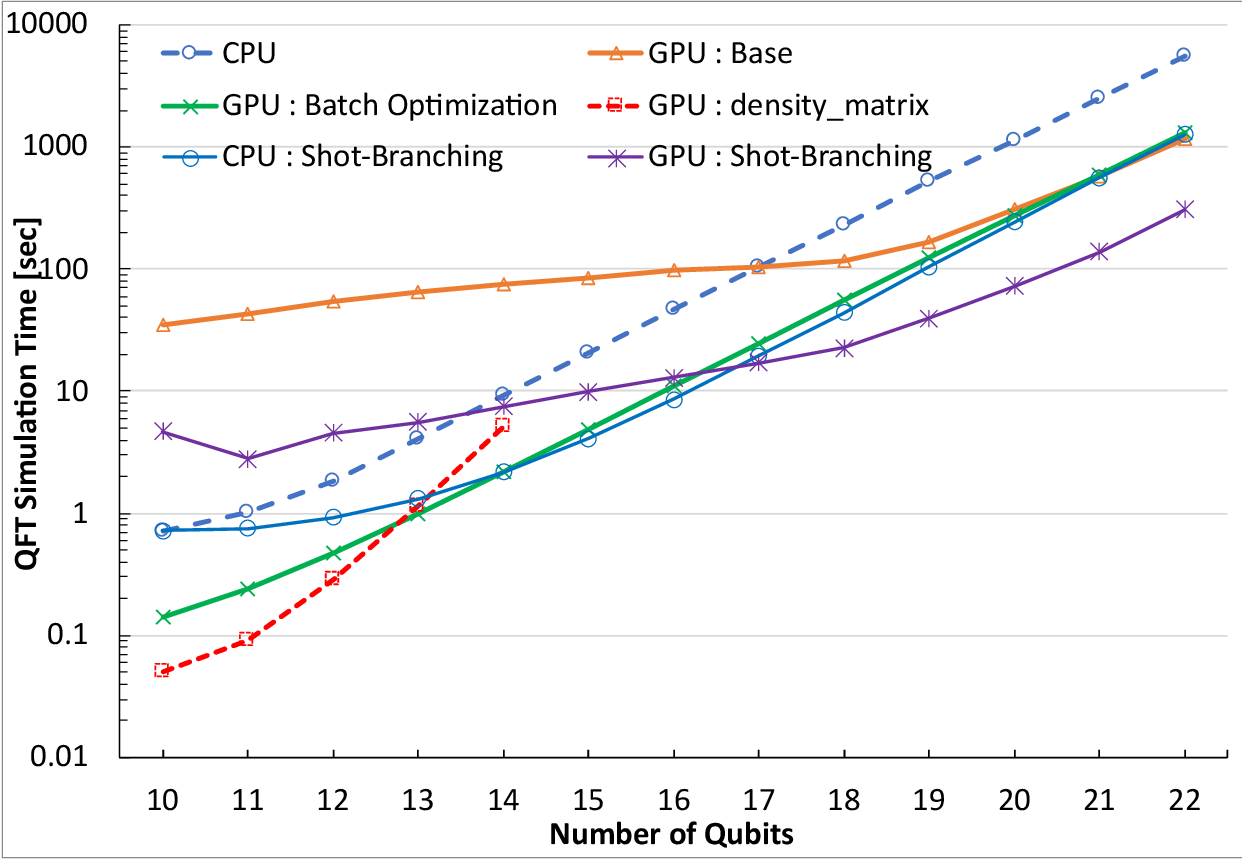} 
\caption{Simulation time comparison of 4,000 shots of QFT circuit with 1\% Kraus noise on Qiskit Aer using a single GPU. (IBM Power System AC922 with 1 NVIDIA Tesla V100.)}
\label{fig_single_GPU_shot_branching_kraus} 
\end{figure}

Fig. \ref{fig_single_GPU_shot_branching_kraus} shows a comparison of the simulation time including the shot-branching technique. We compared the simulation time with and without shot-branching on both the CPU and GPU and with the batch-shots technique on the GPU. By applying shot-branching, we can increase the speed of the noise simulation dramatically. The simulation is slow when the number of qubits is small because of the algorithmic overheads of shot-branching, but these overheads can be ignored easily by increasing the number of qubits. By selecting the technique that shows the best performance, the GPU simulation is now faster than the CPU simulation for all the qubits ranges.

\begin{figure}[!t] \centering \includegraphics[width=2.5in]{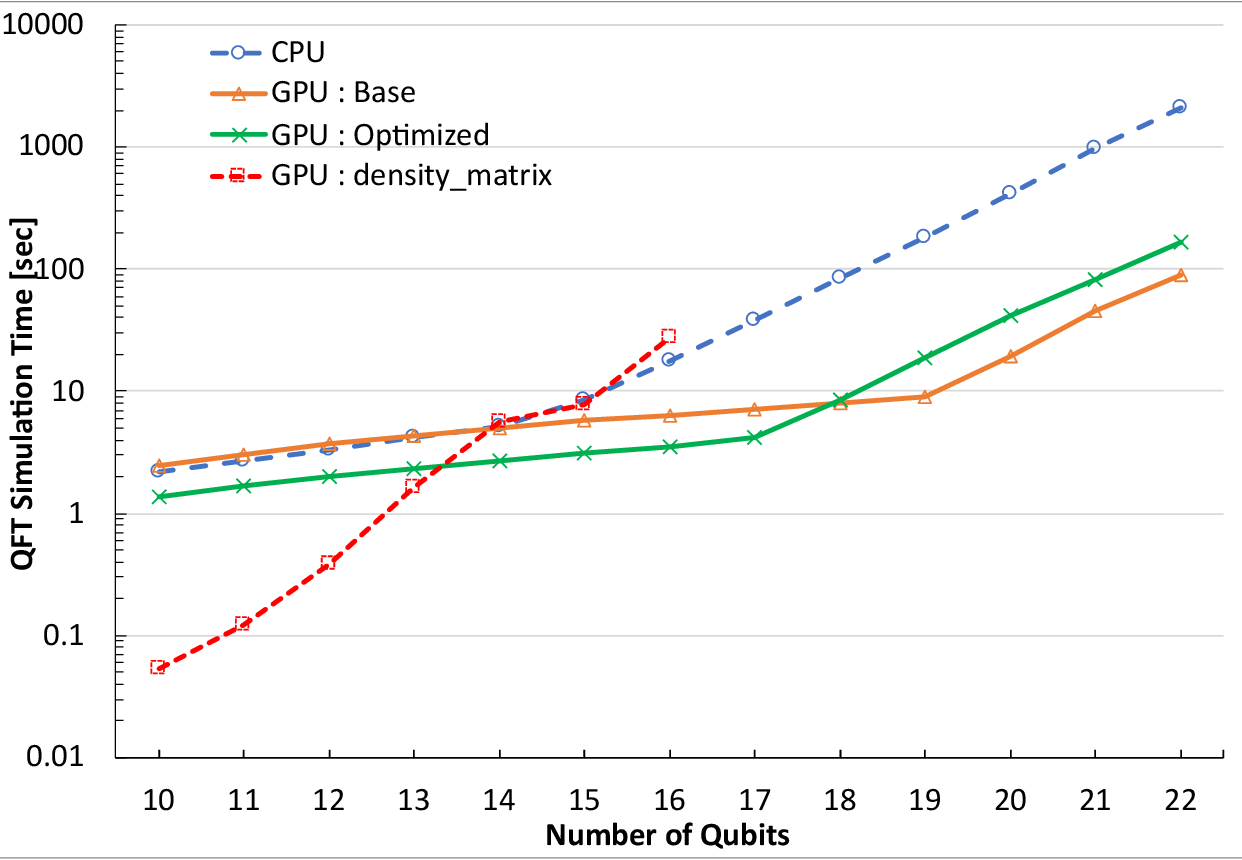} 
\caption{Simulation time comparison of 4,000 shots of QFT circuit with 1\% Pauli noise on Qiskit Aer using six GPUs. (IBM Power System AC922 with six NVIDIA Tesla V100.)}
\label{fig_multi_GPU_pauli} 
\end{figure}

\begin{figure}[!t] \centering \includegraphics[width=2.5in]{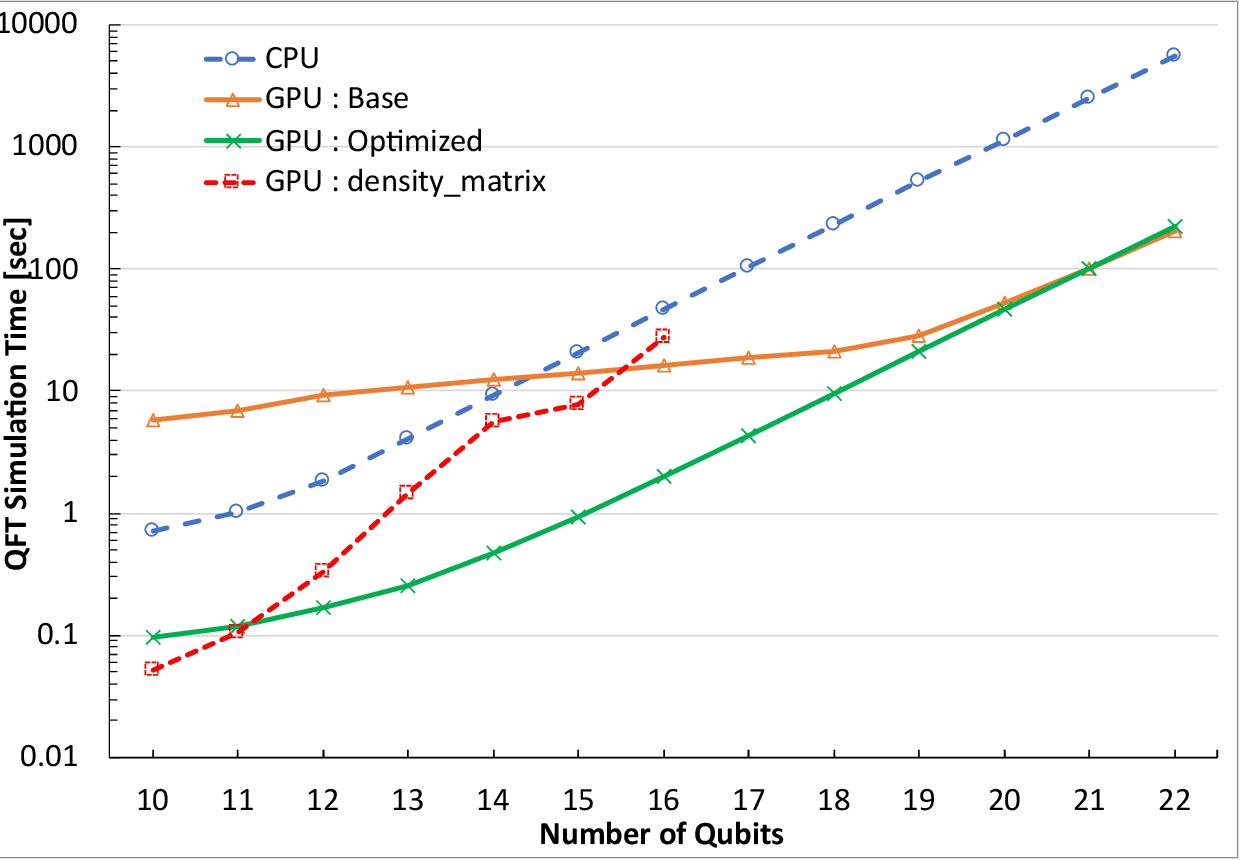} 
\caption{Simulation time comparison of 4,000 shots of QFT circuit with 1\% Kraus noise on Qiskit Aer using six GPUs. (IBM Power System AC922 with six NVIDIA Tesla V100.)}
\label{fig_multi_GPU_kraus} 
\end{figure}

\subsection{Multi-GPU Performance Evaluation}
We compare the simulation time of the QFT circuit with the Pauli and Kraus noise models using all six GPUs and host CPU. In this evaluation, we run 4,000 shots and set the error ratio to 1\%. We also compare the simulation time using the density matrix method using six GPUs.

As shown in Fig. \ref{fig_multi_GPU_pauli}, we compared the simulation time of the QFT circuit with 1\% Pauli noise. Compared to the result for a single GPU, the simulation time for the conventional implementation is shortened by distributing shots on the six GPUs and is faster than running on the CPU. However, the optimized simulation time is not reduced from a single GPU run for smaller qubits because noise sampling on the CPU becomes time dominant since we accelerate the calculation itself. We can obtain the benefits of multiple GPUs from \verb+15-qubits+ or more qubits in this evaluation. The density matrix simulator is parallelized on the six GPUs, so we can increase the maximun number of qubits to 16, but more shots are required to obtain the advantage of the density matrix method. 

As shown in Fig. \ref{fig_multi_GPU_kraus}, we compare the simulation time of the QFT circuit with 1\% Kraus noise. Similarly to the evaluation for Pauli noise, the conventional implementation is well improved by distributing shots on the six GPUs. The simulation time with the batched multi-shot optimization is well distributed on the six GPUs in the Kraus noise case, and there is almost no advantage of the density matrix method for 4,000 shots.

\subsection{Effect of error ratio}
\begin{figure}[!t] \centering \includegraphics[width=2.5in]{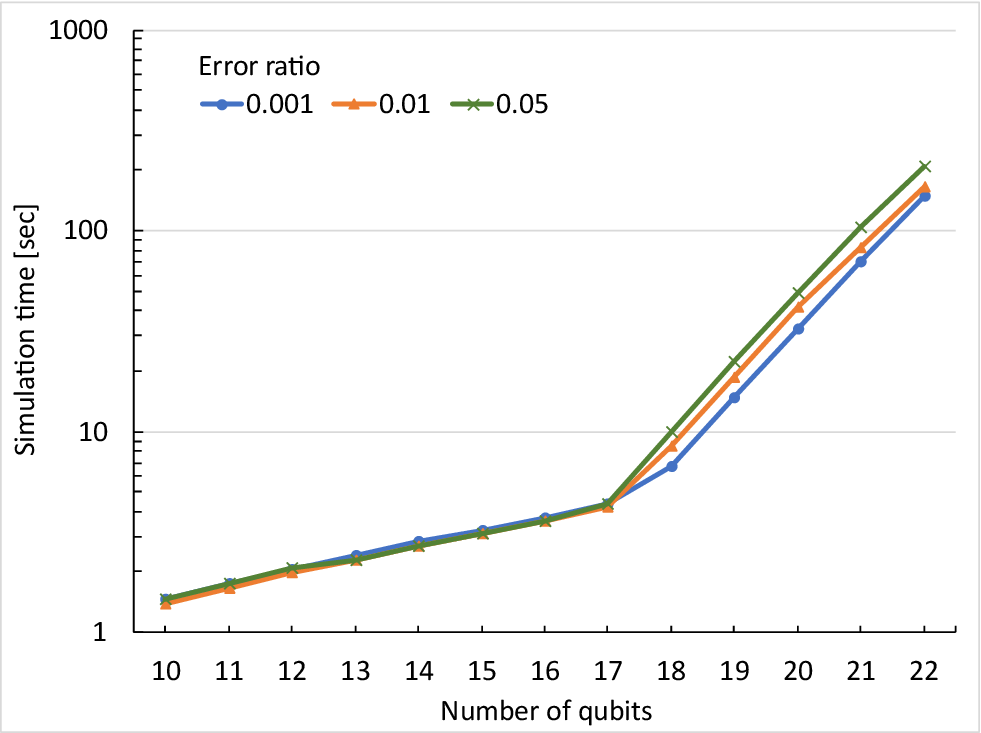} 
\caption{Simulation time comparison changing error ratio in batched multi-shot optimization, 4,000 shots of QFT circuit with 1\% Pauli noise on Qiskit Aer using six GPUs. (IBM Power System AC922 with six NVIDIA Tesla V100.)}
\label{fig_pauli_error_ratio} 
\end{figure}

\begin{figure}[!t] \centering \includegraphics[width=2.5in]{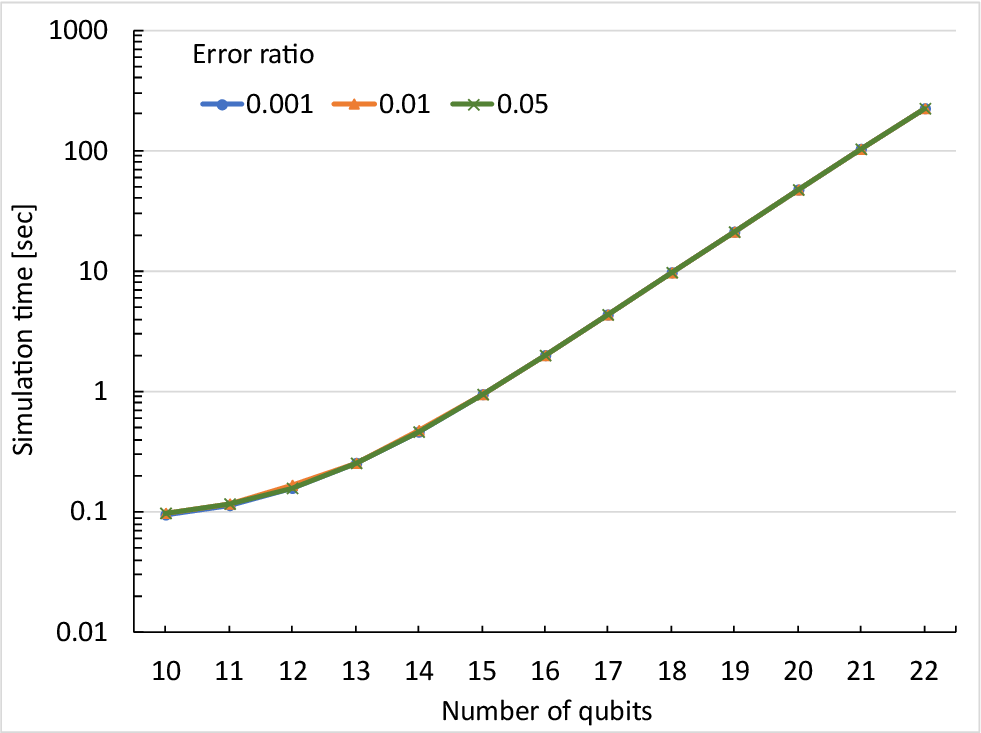} 
\caption{Simulation time comparison changing error ratio in batched multi-shot optimization, 4,000 shots of QFT circuit with 1\% Kraus noise on Qiskit Aer using six GPUs. (IBM Power System AC922 with six NVIDIA Tesla V100.)}
\label{fig_kraus_error_ratio} 
\end{figure}

As shown in Figs. \ref{fig_pauli_error_ratio} and \ref{fig_kraus_error_ratio}, we compared the simulation time by changing the error ratio for each noise model to observe any differences with the batched multi-shot optimization. As shown in Fig. \ref{fig_pauli_error_ratio} there is almost no difference with the lower number of qubits but the simulation time increases as the error ratio increases. We do not need to execute the batched Pauli noise operation when the ID gate is sampled to all shots asthis occurs frequently when the error ratio is small. On the other hand, as shown in Fig. \ref{fig_kraus_error_ratio} the simulation time for the Kraus noise simulation is almost the same compared among the different error ratios because the batched Kraus operator executes all Kraus iterations regardless of the error ratio.

\subsection{Distributing on Multiple Nodes}
We also evaluated the scalability of the simulation time by distributing shots to multiple nodes on the cluster using MPI. We simulate the QFT circuit with the Pauli and Kraus noise models and applied the batched multi-shot optimization technique proposed in this paper. We ran a varied number of shots from 1 to 16 nodes on the IBM Power Sysmte AC922 cluster. 

\begin{figure}[!t] \centering \includegraphics[width=3.0in]{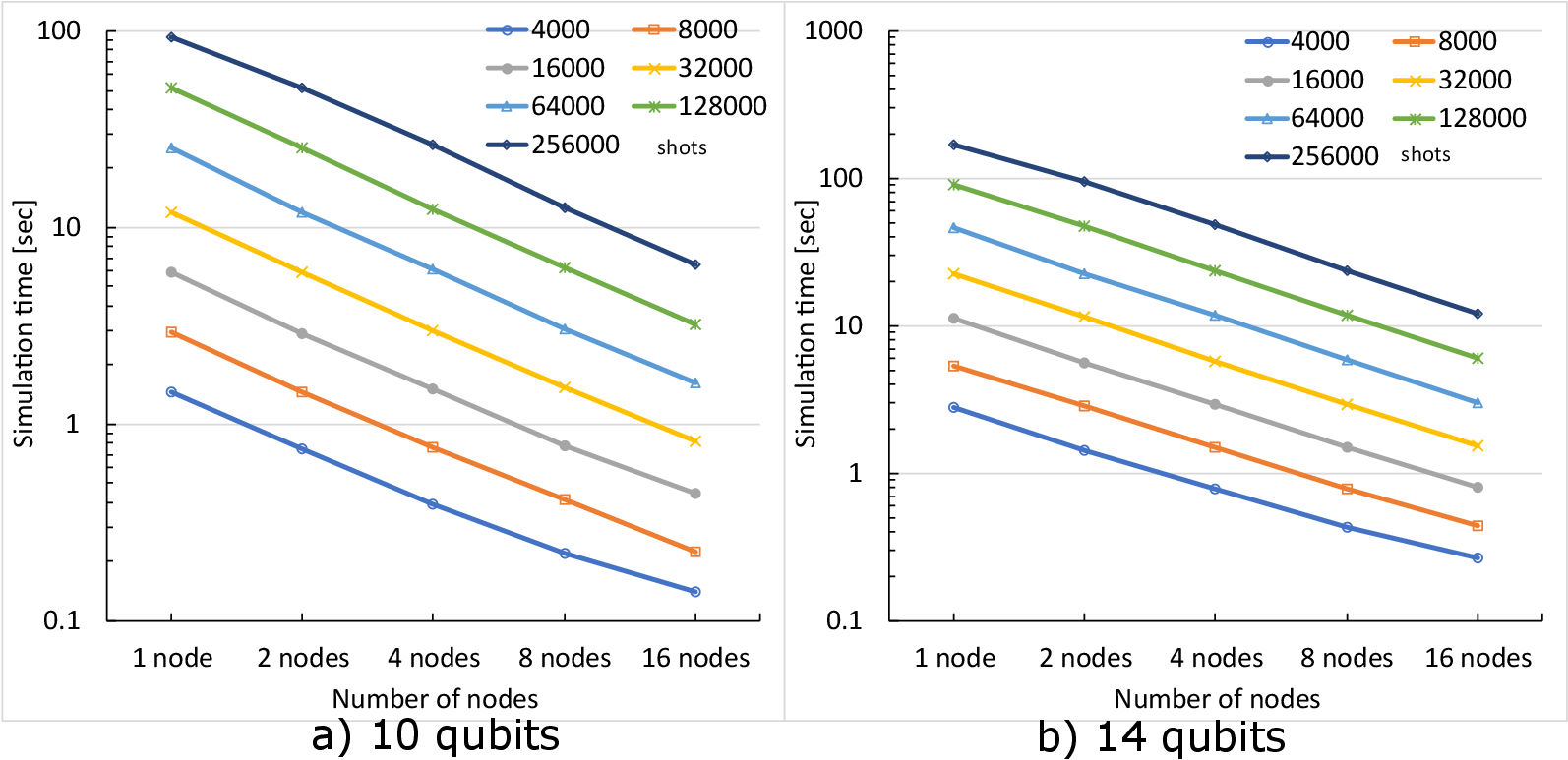} 
\caption{Multi-node simulation time comparison of QFT circuit with 1\% Pauli noise on Qiskit Aer. (IBM Power System AC922 cluster with six NVIDIA Tesla V100.)}
\label{fig_MPI_pauli} 
\end{figure}

\begin{figure}[!t] \centering \includegraphics[width=3.0in]{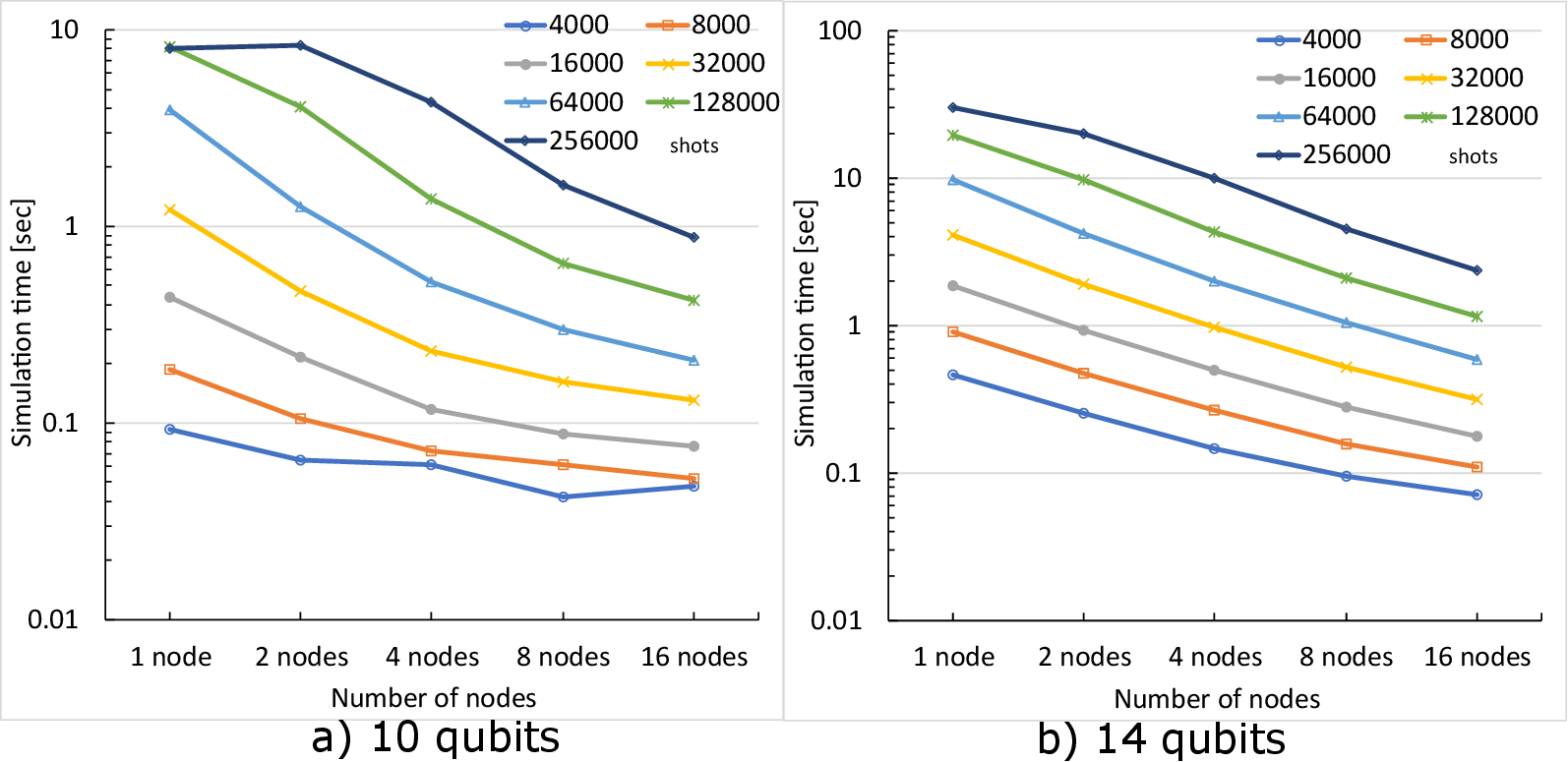} 
\caption{Multi-node simulation time comparison of QFT circuit with 1\% Kraus noise on Qiskit Aer. (IBM Power System AC922 cluster with six NVIDIA Tesla V100.)}
\label{fig_MPI_kraus} 
\end{figure}

Figs. \ref{fig_MPI_pauli} and. \ref{fig_MPI_kraus} show the scalabilities of the noise simulations on the cluster. Comparing the two noise models, the Pauli noise simulation scales better than the Kraus noise one because it requires more time. Since the \verb+10-qubits+ simulation with Kraus noise requires less time, scalability is ineffective because of the distribution and parallelization overheads. To improve scaling on the cluster, we require sufficient qubits and shots, but our optimization technique shows very good scalability on the cluster.

\section{Summary}
Simulating quantum circuits with noise models is very important for the development of quantum computers and algorithms on classical computers, but it takes a long time to simulate multiple shots of simulations. GPUs can contribute to the development of quantum computers by accelerating simulations, but they are ineffective at accelerating circuits with a small number of qubits because of overheads and noise simulation randomness.

In this paper, we introduced an optimization technique to batch multi-shots running on a single GPU kernel program to decrease overheads loaded on the host CPU for a small number of qubits. We applied a batched multi-shots technique to simulate Pauli and Kraus noises and implemented it in an open-source quantum computing simulation platform Qiskit Aer. 

The batched multi-shots optimization decreases GPU overheads but increases unnecessary computation and memory usage. This strategy is in contrast with the shot-branching technique introduced in this paper, which reduces computation by sharing states between shots. The shot-branching technique shows very good performance both on the CPU and GPU when the number of qubits is sufficiently large.

The conventional multi-shot simulations with noise models were slower than the simulations on the CPU for circuits with a small number of qubits. By applying the batched multi-shot optimization, simulating on the GPU is now faster than the CPU over the entire qubits range, and by applying shot-branching, we improve performance more and finally we increase the performance speed ten to a hundred times compared with that of the conventional implementation. 

Our techniques to optimize multi-shot simulations cover the entire qubits range; the batched multi-shot technique is suitable for a smaller number of qubits, and the shot-branching technique shows faster speeds for a larger number of qubits. However, selecting the simulation method and implementation suitable for the actual applications is necessary. 

We also distributed multi-shot simulations to multiple GPUs and multiple compute nodes on clusters using MPI. We obtained good scaling on the cluster to simulate thousands of shots. 

This technique can also be applied to simulating multiple shots with varied parameters in a number of gates (i.e. theta for rotational gates) or multiple circuits. We will implement GPU acceleration support for various multi-shot and multi-circuit simulations in the future.

We believe that we can combine batched multi-shots optimization and shot-branching. We will implement the combined implementation to further improve the speed of multi-shot simulations.

We are focusing on accelerating multi-shot simulations on a GPU; however, the same technique can be applied to accelerate simulations on a CPU. The batched multi-shot optimization has the potential to utilize SIMD instructions efficiently because it increases the vector length to be calculated. 

Our batched multi-shots optimization implemented in Qiskit Aer does not use the cuStateVec APIs of cuQuantum SDK currently because cuStateVec lacks the means to mask the calculation for each shot by referring to classical bits. We will exploit cuStateVec to accelerate multi-shot simulations by applying a technique suitable for cuStateVec APIs.

\bibliographystyle{IEEEtran} \bibliography{IEEEabrv,mybib}

\begin{thebibliography}{10}
\providecommand{\url}[1]{#1}
\csname url@samestyle\endcsname
\providecommand{\newblock}{\relax}
\providecommand{\bibinfo}[2]{#2}
\providecommand{\BIBentrySTDinterwordspacing}{\spaceskip=0pt\relax}
\providecommand{\BIBentryALTinterwordstretchfactor}{4}
\providecommand{\BIBentryALTinterwordspacing}{\spaceskip=\fontdimen2\font plus
\BIBentryALTinterwordstretchfactor\fontdimen3\font minus
  \fontdimen4\font\relax}
\providecommand{\BIBforeignlanguage}[2]{{%
\expandafter\ifx\csname l@#1\endcsname\relax
\typeout{** WARNING: IEEEtran.bst: No hyphenation pattern has been}%
\typeout{** loaded for the language `#1'. Using the pattern for}%
\typeout{** the default language instead.}%
\else
\language=\csname l@#1\endcsname
\fi
#2}}
\providecommand{\BIBdecl}{\relax}
\BIBdecl

\bibitem{qiskit-aer}
\BIBentryALTinterwordspacing
{Qiskit Aer: A High Performance Simulator Framework for Quantum Circuits}.
  [Online]. Available: \url{https://qiskit.org/aer}
\BIBentrySTDinterwordspacing

\bibitem{qsim}
\BIBentryALTinterwordspacing
qsim optimized quantum circuit simulators. [Online]. Available:
  \url{https://quantumai.google/qsim}
\BIBentrySTDinterwordspacing

\bibitem{cirq}
\BIBentryALTinterwordspacing
Cirq an open source framework for programming quantum computers. [Online].
  Available: \url{https://quantumai.google/cirq}
\BIBentrySTDinterwordspacing

\bibitem{Suzuki2021qulacsfast}
\BIBentryALTinterwordspacing
Y.~Suzuki, Y.~Kawase, Y.~Masumura, Y.~Hiraga, M.~Nakadai, J.~Chen, K.~M.
  Nakanishi, K.~Mitarai, R.~Imai, S.~Tamiya, T.~Yamamoto, T.~Yan, T.~Kawakubo,
  Y.~O. Nakagawa, Y.~Ibe, Y.~Zhang, H.~Yamashita, H.~Yoshimura, A.~Hayashi, and
  K.~Fujii, ``Qulacs: a fast and versatile quantum circuit simulator for
  research purpose,'' \emph{{Quantum}}, vol.~5, p. 559, Oct. 2021. [Online].
  Available: \url{https://doi.org/10.22331/q-2021-10-06-559}
\BIBentrySTDinterwordspacing

\bibitem{efthymiou2021qibo}
S.~Efthymiou, S.~Ramos-Calderer, C.~Bravo-Prieto, A.~P{\'e}rez-Salinas,
  D.~Garc{\'\i}a-Mart{\'\i}n, A.~Garcia-Saez, J.~I. Latorre, and S.~Carrazza,
  ``Qibo: a framework for quantum simulation with hardware acceleration,''
  \emph{Quantum Science and Technology}, vol.~7, no.~1, p. 015018, 2021.

\bibitem{jones2018quest}
T.~Jones, A.~Brown, I.~Bush, and S.~Benjamin, ``Quest and high performance
  simulation of quantum computers,'' \emph{arXiv preprint arXiv:1802.08032},
  2018.

\bibitem{li2017quantum}
Z.~Li and J.~Yuan, ``Quantum computer simulation on gpu cluster incorporating
  data locality,'' in \emph{International Conference on Cloud Computing and
  Security}.\hskip 1em plus 0.5em minus 0.4em\relax Springer, 2017, pp. 85--97.

\bibitem{haner20170}
T.~H{\"a}ner and D.~S. Steiger, ``0.5 petabyte simulation of a 45-qubit quantum
  circuit,'' in \emph{Proceedings of the International Conference for High
  Performance Computing, Networking, Storage and Analysis}.\hskip 1em plus
  0.5em minus 0.4em\relax ACM, 2017, p.~33.

\bibitem{smelyanskiy2016qhipster}
M.~Smelyanskiy, N.~P. Sawaya, and A.~Aspuru-Guzik, ``qhipster: the quantum high
  performance software testing environment,'' \emph{arXiv preprint
  arXiv:1601.07195}, 2016.

\bibitem{DeRaedt2019}
H.~Raedt, D.~Willsch, M.~Nocon, N.~Yoshioka, N.~Ito, S.~Yuan, and
  K.~Michielsen, ``Massively parallel quantum computer simulator, eleven years
  later,'' \emph{Computer Physics Communications}, vol. 237, 11 2018.

\bibitem{wu2019}
\BIBentryALTinterwordspacing
X.-C. Wu, S.~Di, E.~M. Dasgupta, F.~Cappello, H.~Finkel, Y.~Alexeev, and F.~T.
  Chong, ``Full-state quantum circuit simulation by using data compression,''
  in \emph{Proceedings of the International Conference for High Performance
  Computing, Networking, Storage and Analysis}, ser. SC ’19.\hskip 1em plus
  0.5em minus 0.4em\relax New York, NY, USA: Association for Computing
  Machinery, 2019. [Online]. Available:
  \url{https://doi.org/10.1145/3295500.3356155}
\BIBentrySTDinterwordspacing

\bibitem{doi2019}
\BIBentryALTinterwordspacing
J.~Doi, H.~Takahashi, R.~Raymond, T.~Imamichi, and H.~Horii, ``Quantum
  computing simulator on a heterogenous hpc system,'' in \emph{Proceedings of
  the 16th ACM International Conference on Computing Frontiers}, ser. CF
  ’19.\hskip 1em plus 0.5em minus 0.4em\relax New York, NY, USA: Association
  for Computing Machinery, 2019, pp. 85--93. [Online]. Available:
  \url{https://doi.org/10.1145/3310273.3323053}
\BIBentrySTDinterwordspacing

\bibitem{doi2020}
J.~Doi and H.~Horii, ``Cache blocking technique to large scale quantum
  computing simulation on supercomputers,'' in \emph{2020 IEEE International
  Conference on Quantum Computing and Engineering (QCE)}, 2020, pp. 212--222.

\bibitem{Liu2021}
\BIBentryALTinterwordspacing
Y.~A. Liu, X.~L. Liu, F.~N. Li, H.~Fu, Y.~Yang, J.~Song, P.~Zhao, Z.~Wang,
  D.~Peng, H.~Chen, C.~Guo, H.~Huang, W.~Wu, and D.~Chen, ``Closing the
  "quantum supremacy" gap: Achieving real-time simulation of a random quantum
  circuit using a new sunway supercomputer,'' in \emph{Proceedings of the
  International Conference for High Performance Computing, Networking, Storage
  and Analysis}, ser. SC '21.\hskip 1em plus 0.5em minus 0.4em\relax New York,
  NY, USA: Association for Computing Machinery, 2021. [Online]. Available:
  \url{https://doi.org/10.1145/3458817.3487399}
\BIBentrySTDinterwordspacing

\bibitem{isakov2021}
\BIBentryALTinterwordspacing
S.~V. Isakov, D.~Kafri, O.~Martin, C.~V. Heidweiller, W.~Mruczkiewicz, M.~P.
  Harrigan, N.~C. Rubin, R.~Thomson, M.~Broughton, K.~Kissell, E.~Peters,
  E.~Gustafson, A.~C.~Y. Li, H.~Lamm, G.~Perdue, A.~K. Ho, D.~Strain, and
  S.~Boixo, ``Simulations of quantum circuits with approximate noise using qsim
  and cirq,'' 2021. [Online]. Available:
  \url{https://doi.org/10.48550/arxiv.2111.02396}
\BIBentrySTDinterwordspacing

\bibitem{benenti2007}
C.~G. . S.~G. Benenti, G., \emph{Principles of Quantum Computation and
  Information-Volume II: Basic Tools and Special Topics}.\hskip 1em plus 0.5em
  minus 0.4em\relax Wspc; Illustrated, 2007.

\bibitem{amazon_braket}
\BIBentryALTinterwordspacing
Amazon braket accelerate quantum computing research. [Online]. Available:
  \url{https://aws.amazon.com/braket}
\BIBentrySTDinterwordspacing

\bibitem{Zulehner2017}
\BIBentryALTinterwordspacing
A.~Zulehner and R.~Wille, ``Advanced simulation of quantum computations,''
  2017. [Online]. Available: \url{https://arxiv.org/abs/1707.00865}
\BIBentrySTDinterwordspacing

\bibitem{Li2019}
\BIBentryALTinterwordspacing
G.~Li, Y.~Ding, and Y.~Xie, ``Sanq: A simulation framework for architecting
  noisy intermediate-scale quantum computing system,'' 2019. [Online].
  Available: \url{https://arxiv.org/abs/1904.11590}
\BIBentrySTDinterwordspacing

\bibitem{Wang2022}
\BIBentryALTinterwordspacing
M.~Wang, R.~Huang, S.~Tannu, and P.~Nair, ``Tqsim: A case for reuse-focused
  tree-based quantum circuit simulation,'' 2022. [Online]. Available:
  \url{https://arxiv.org/abs/2203.13892}
\BIBentrySTDinterwordspacing

\bibitem{qiskit}
\BIBentryALTinterwordspacing
Qiskit: An open-source framework for quantum computing. [Online]. Available:
  \url{https://qiskit.org/}
\BIBentrySTDinterwordspacing

\bibitem{Thrust}
\BIBentryALTinterwordspacing
Thrust - parallel algorithms library. [Online]. Available:
  \url{https://thrust.github.io}
\BIBentrySTDinterwordspacing

\bibitem{cuquantum}
\BIBentryALTinterwordspacing
cuquantum accelerate quantum computing research. [Online]. Available:
  \url{https://developer.nvidia.com/cuquantum-sdk}
\BIBentrySTDinterwordspacing

\bibitem{custatevec}
\BIBentryALTinterwordspacing
custatevec: A high-performance library for state vector quantum simulators.
  [Online]. Available:
  \url{https://docs.nvidia.com/cuda/cuquantum/custatevec/index.html}
\BIBentrySTDinterwordspacing

\bibitem{caldeira2018ibm}
A.~Caldeira, ``Ibm power system ac922 introduction and technical overview,''
  \emph{IBM Redbooks}, 2018.

\bibitem{tesla-v100}
\BIBentryALTinterwordspacing
Nvidia tesla v100. [Online]. Available:
  \url{https://www.nvidia.com/en-us/data-center/tesla-v100/}
\BIBentrySTDinterwordspacing

\bibitem{nvlink}
\BIBentryALTinterwordspacing
Nvlink interconnect technology. [Online]. Available:
  \url{http://www.nvidia.com/object/nvlink.html}
\BIBentrySTDinterwordspacing

\end{thebibliography}

\end{document}